\documentclass[%
reprint,
 amsmath,amssymb,
 aps,
floatfix,
]{revtex4-1}

\usepackage{graphicx}
\usepackage{dcolumn}
\usepackage{bm}
\usepackage{soul}
\usepackage[usenames, dvipsnames]{color}
\usepackage{mathtools}

\DeclarePairedDelimiter\ket{\lvert}{\rangle}
\usepackage[bordercolor=gray!20,backgroundcolor=orange!10,linecolor=none,textsize=footnotesize,textwidth=0.6in]{todonotes}
\begin{document}

\preprint{APS/123-QED}

\title{Optimizing the formation of depth-confined nitrogen vacancy center spin ensembles in diamond for quantum sensing}

\author{Tim R. Eichhorn}
\altaffiliation[These authors contributed equally to this work ]{}
\author{Claire A. McLellan}%
\altaffiliation[These authors contributed equally to this work ]{}
\author{Ania C. Bleszynski Jayich}
 \email{ania@physics.ucsb.edu}
\affiliation{%
 University of California, Santa Barbara\\
 Physics Department
}%

\date{\today}

\begin{abstract}


Spin ensembles of nitrogen vacancy (NV) centers in diamond are emerging as powerful spin-based sensors for magnetic, electric and thermal field imaging with high spatial and temporal resolution. Here we characterize the formation of depth-confined NV center ensembles, activated by electron irradiation in diamond layers grown by plasma enhanced chemical vapor deposition with nitrogen co-doping. To do so, we exploit the high magnetic sensitivity of ensembles of NV centers to probe their spin environment as a function of growth and irradiation parameters. We engineer an NV ensemble whose magnetic sensitivity is within a factor of two of the static NV-NV dipolar interaction limit, thus demonstrating a powerful platform for quantum sensing.

\end{abstract}

\maketitle

\section{\label{sec:intro}Introduction}
Nitrogen vacancy (NV) centers in diamond are excellent sensors of magnetic and electric fields, temperature, and strain due to their long quantum coherence times and simple optical addressability \cite{doherty2013nitrogen}. While single NV centers can exhibit excellent spatial resolution \cite{pelliccione2016scanned,maletinsky2012robust,thiel2016quantitative}, scanning a single NV center over a large area for imaging is inherently slow. Ensembles of NV center spins, localized close to the diamond surface, present a powerful platform for imaging mesoscopic phenomena with high sensitivity and high spatiotemporal resolution when operated in a widefield imaging modality. Recently, NV ensembles have been used to image magnetic fields in biological and condensed matter systems, achieving ms-scale temporal resolution \cite{barry2016optical} and micron-scale spatial resolution \cite{tetienne2018spin,simpson2017electron}. 

While ensembles of NV centers have reached sub-picoTesla magnetic sensitivity in millimeter-sized sensing volumes \cite{wolf2015subpicotesla}, increasing the sensitivity of ensembles of NV centers in small sensing volumes is an outstanding challenge. The sensitivity $\eta$ of a spin ensemble scales with the square root of its quantum coherence time, $T_2$, and the total number of NV centers, $N_{\text{NV}}$: $\eta\propto\frac{1}{\sqrt[]{N_{\text{NV}}\cdot T_2}}$, and hence it is desirable to have as many sensor spins as possible in a small volume while maintaining their long coherence time \cite{acosta2009diamonds}. As spin-spin interactions contribute to decoherence, the targeted density of NV sensor spins should be high compared to all other paramagnetic defects, such as P1 centers and vacancy related defects, so that mutual NV spin interactions are the dominant source of decoherence. Decoherence due to, \textit{e.g.}, strain inhomogeneities and $^{13}$C nuclear spins must also be minimized. Though the decohering effects of a non-NV spin bath can be mitigated via driving techniques, as demonstrated in a recent experiment \cite{bauch2018ultralong}, spin bath driving involves much added complexity. A starting material in which NV coherence is limited by the NV-spin bath itself provides a powerful platform for sensing experiments and for implementing techniques that harness quantum correlations to further enhance sensitivity \cite{cappellaro2009quantum,choi2017quantum,Zhou2018}.

Pathways towards creating NV center-rich diamond all require nitrogen-rich diamond, a means of generating a high density of vacancies, and subsequent annealing to form NV centers. A high nitrogen concentration can be realized in high pressure high temperature (HPHT) diamond, whose synthesis results in high nitrogen content \cite{palyanov2010effect}, diamond implanted with nitrogen ions, or PECVD-grown diamond doped with nitrogen during growth. For high-spatial resolution sensing, HPHT diamond has the disadvantage that nitrogen is uniformly distributed throughout the diamond, making it difficult to remove signal from NV centers far from the surface of the diamond, increasing the background counts and therefore diminishing sensitivity and spatial resolution. Nitrogen ion implantation creates collateral damage in the lattice that can degrade the coherence of individual NV spins \cite{naydenov2010increasing}. Nitrogen-doping during plasma enhanced chemical vapor deposition (PECVD) growth of diamond has produced reproducibly highly coherent single NV centers with nanometer-scale depth control\cite{ohno2012,mclellan2016}, which has intriguing prospects for ensemble-based sensing. However, despite the promise, little work has focused on generating and characterizing high-density NV spin ensembles formed via PECVD growth with depth-confined nitrogen doping.

In this paper we combine nitrogen-doped, PECVD-grown diamond with electron irradiation from a transmission electron microscope to realize high-density ensembles of NV sensor spins localized within a few hundred nanometers of the surface. To optimize the tailor-made diamond, we tune the nitrogen incorporation during growth as well as the vacancy creation process via electron irradiation dosage and energy.  To facilitate a wide exploration of NV formation parameter space, we utilize an important advantage of the TEM irradiation technique: the ability to confine electron irradiation to micrometer-scale diameter spots, which allows for a large range of energies and dosages across a single sample. We subsequently characterize the diamond's spin environment and its interactions with the NV centers using double electron-electron resonance (DEER) methods and instantaneous diffusion. We identify decoherence stemming from P1 centers and NV centers, with relative strengths tuned by the parameters of the NV formation process. We hone in on a range of NV formation parameters where the NV ensemble coherence is limited by static NV-NV dipolar interactions and the formation of other defects is negligible. Decoherence due to other sources, such as strain and $^{13}$C nuclear spins, is also negligible in our tailor-made diamond. 

\begin{figure}
\includegraphics[width=\columnwidth]{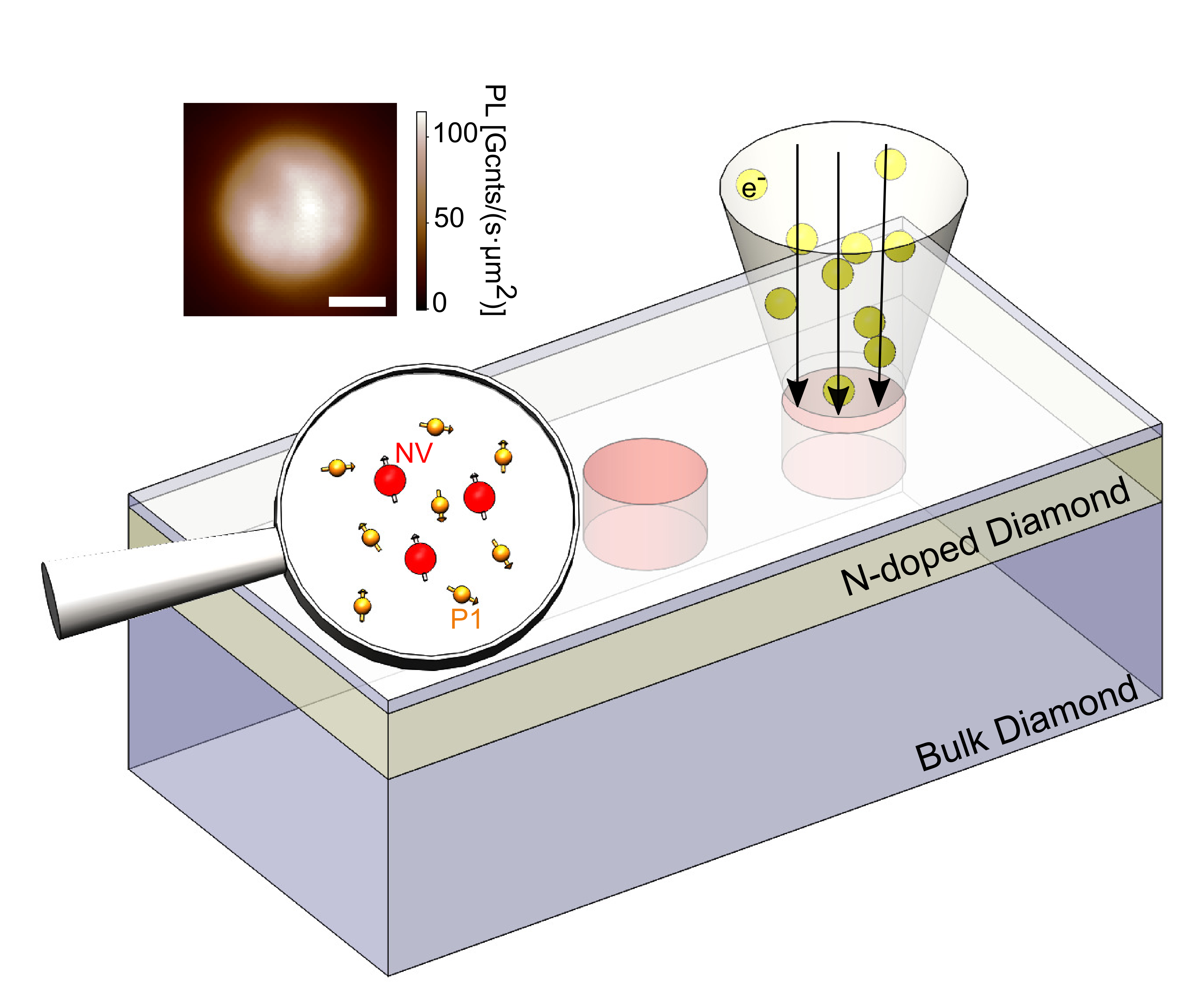}
\caption{Schematic of a shallow, nitrogen-rich diamond layer irradiated with electrons in a series of 5-10 $\mu m$ diameter spots (indicated by pink circles) across the diamond. Magnifying glass highlights the P1 and NV center spins introduced in the diamond lattice. Inset is a fluorescence image of a TEM-irradiated spot upon 145 keV electron irradiation and 10$^{21}$ e$^-$/cm$^2$ dosage.  We note that the spatially nonuniform fluorescence within the spot is a result of an inhomogeneous laser intensity profile.  Scale bar is 2 $\mu$m.}
\label{fig:fig1}
\end{figure}


\section{\label{sec:diamond_fab}NV ensembles in PECVD-grown, electron-irradiated diamonds}
\subsection{\label{subsec:diamond_chips}Sample preparation}
\begin{table*}
\centering
\begin{ruledtabular}
\begin{tabular}{lllll}
Sample name & Nitrogen flow rate & Nitrogen isotope & Electron energies & Electron dose range\\
 C031 & 0.1 sccm  &  $^{15}$N &  145 keV & $6.3\cdot10^{20} - 2.5\cdot10^{21} $e$^-$/cm$^2$ \\
 C041 & 1 sccm  &  $^{15}$N & 145 and 200 keV &  $10^{19} - 10^{22}$ e$^-$/cm$^2$ \\
 C044 & 5 sccm &  $^{14}$N & 145 and 200 keV & $10^{19} - 10^{22}$ e$^-$/cm$^2$ 
\end{tabular}
\end{ruledtabular}
\caption{Parameters for diamond samples used in this work}
\label{tab:parameterTable}
\end{table*}

The single-crystal diamond films studied here are fabricated in house by nitrogen doping during plasma enhanced chemical vapor deposition followed by electron irradiation and annealing \cite{ohno2012,mclellan2016}. The samples are grown on commercially available 2x2 mm$^2$ electronic grade diamond substrates (Element Six). Prior to growth the samples are polished to sub-nanometer surface roughness and etched 500 nm using an Ar/Cl inductively coupled plasma to remove polishing-induced strain. Three samples are studied here: C031, C041, and C044.  

C031 and C041 are doped with 99\% $^{15}$N isotopically pure gas and C044 is grown with natural (99\% $^{14}$N) isotopic purity gas. The growth proceeds as follows: a 32-nm undoped diamond buffer layer is grown, followed by a 500-nm thick nitrogen-doped layer formed by introducing nitrogen gas with a flowrate of 0.1 - 5 sccm. The nitrogen is then turned off and a final 50-nm diamond cap is grown. The samples are then irradiated by a transmission electron microscope with 145 or 200 keV electrons in doses ranging from $10^{19} - 10^{22}$  e$^-$/cm$^2$. The electron irradiation is done in 10-$\mu$m diameter spots on sample C031 and 5-$\mu $m diameter spots on samples C041 and C044. After irradiation the samples are annealed for 48 hours in Ar/H$_2$ forming gas at $850\,^\circ$C to activate vacancy diffusion. Samples C041 and C044 are then cleaned in boiling mixture of  nitric, sulfuric, and perchloric  acids (1:1:1 mixture ratio) for 1 hour; C031, being only 20 $\mu$m in thickness, was considered too fragile for this process. All samples are then annealed for 4 hours in an oxygen atmosphere at $450\,^\circ$C. More details on the PECVD growth and NV formation can be found in the supplement.

Figure \ref{fig:fig1}a is a schematic of the samples formed in this work, showing a shallow, nitrogen-rich diamond layer irradiated locally with electrons in 5-10 $\mu$m diameter spots, where the electron dosage and energy vary between spots. In each spot the dominant spin species, P1 centers and NV centers, are highlighted. The inset shows a confocal image of a typical TEM-irradiated spot, where the bright fluorescence ($\sim 10^{11}$ counts$/$(s$\cdot \mu\text{m}^2)$) at only few percent optical saturation stems from the high density of NV centers formed. Table \ref{tab:parameterTable} presents the three samples used here and lists the relevant parameters that vary between samples. 

\subsection{\label{subsec:equipment} Methods}

Measurements are taken on a homebuilt wide-field microscope under ambient conditions. A 520-nm diode laser excites the NV center ensembles and imaging is performed with a charge coupled device (CCD) camera. For data presented in which pulse sequences are used, NV center fluorescence from the entire excitation spot is focused onto a avalanche photo diode (APD), with sufficient attenuation to avoid saturation of the APD.  Radiofrequency (RF) signal generators are used for controlling the NV centers and P1 centers. Mulitple RF signals ($\sim$ GHz for NVs and 100's of MHz for P1 centers) are combined before amplification and sent through a common RF antenna fabricated on a glass cover-slip, on which the diamond is placed. A circular antenna geometry is chosen to reduce inhomogeneities in microwave power across the excitation area on the diamond.

\section{\label{sec:measurements}Measurement techniques}

Carefully quantifying both the P1 precursor spin density ($n_{\text{P1}}$) and the NV density ($n_{\text{NV}})$ are central to this work. The P1 spin density ultimately limits the NV density and is a source of spin decoherence. As it is difficult to measure the small number of spins present in our samples with standard bulk EPR techniques, whose spin number sensitivities are typically $\sim 10^{10}$ per 1 G linewidth \cite{Eaton2010} at room temperature and X-band, we utilize the high sensitivity of the NV center spin in conjunction with double electron - electron resonance (DEER) techniques to probe $n_{\text{P1}}$ and $n_{\text{NV}}$. Specifically, we use the NV spin ensemble to detect its average magnetic field environment, consisting of other NV spins, residual P1 precursor spins, and any other sources such as vacancy related defects.  We use Hahn echo-based DEER measurements as well as instantaneous diffusion effects to measure the spin bath densities by quantifying the interactions in the different spin baths. 

\subsection{\label{subsec:DEER-sequence}Hahn echo-based double electron electron resonance (DEER) sensing}

\begin{figure}
\includegraphics[width=\columnwidth]{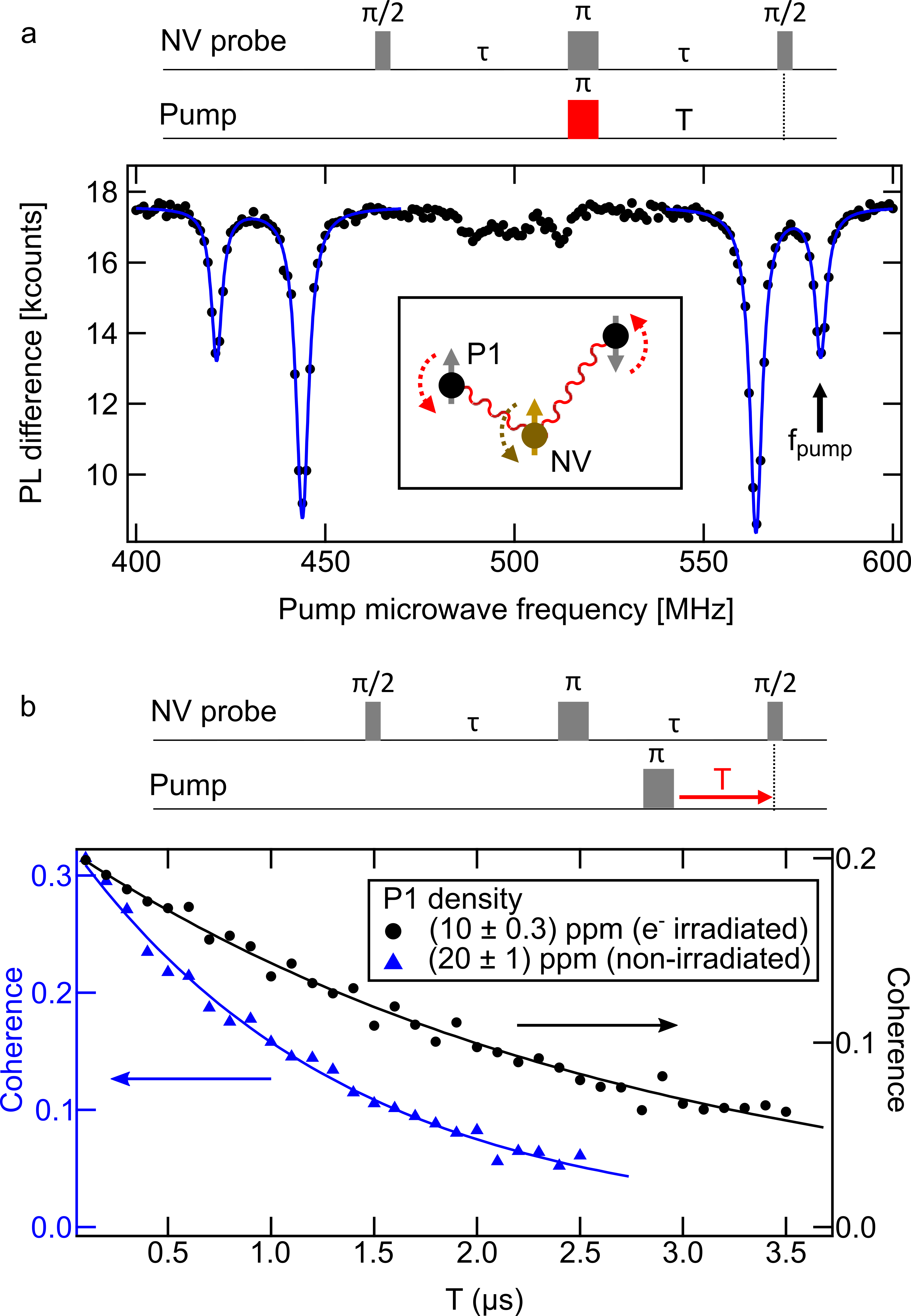}
\caption{P1 spectrum and spin density. (a) Frequency-swept DEER spectrum of P1 centers in a 180 G external magnetic field with fits to the dominant P1 electronic spin transitions, showing the characteristic P1 hyperfine coupling. The pulse sequence is shown: the delay time between the NV probe pulses, $\tau$, is fixed to approximately half the $T_2$ of the NV ($T_2$ = 3.8 $\mu s$ here) and the time between the pump $\pi$ pulse and final probe $\pi/2$ pulse, T, is fixed. The pump and probe $\pi$ pulses are offset by 50 ns to protect the microwave amplifier. Data taken on a non-irradiated area of sample C041. (b) Hahn echo coherence decay taken with the pump-pulse-swept DEER sequence shown (pump frequency is fixed to f$_{\text{pump}}$). Solid lines are mono-exponential fits that yield the P1 density (Eq. \ref{eq:DEER_decay}). Data taken on sample C041 on a non-irradiated area ($n_\text{P1}$ = 20 ppm) and on a spot irradiated with 200 keV electrons at a dose of 10$^{20}$ e$^-$/cm$^2$ ($n_\text{P1}$ = 10 ppm).
}

\label{fig:DEER_Hahn_echo}
\end{figure}

 Figure \ref{fig:DEER_Hahn_echo} shows the Hahn echo-based DEER measurements used to probe spin bath densities. Here, the NV centers are the probe spins; when subject to a Hahn echo pulse sequence these probe spins respond maximally to a static magnetic environment that is inverted by a pump $\pi$-pulse. This pump $\pi$-pulse recouples the static dipolar interaction between the bath and NV spins. 
We first identify the P1 spin transition frequencies as seen in Fig. \ref{fig:DEER_Hahn_echo}a, which shows a P1 center spectrum taken in an external magnetic field ($B_0$) of 180 G aligned to the (111) crystal axis. The spectrum, obtained by sweeping the frequency of the pump (P1) $\pi$-pulse and monitoring the NV center fluorescence, shows the characteristic $^{15}$N P1 center spectrum \cite{cox199413c}. 

When the pump $\pi$ pulse is resonant with a P1 transition, the dipolar interactions between NV and P1 spins are recoupled, causing a drop in the coherence of the NV center probe, which is manifest as a drop in NV photoluminesence (PL) signal. The PL signal that is plotted is the difference between the $\ket {m_s=0}$ and $\ket {m_s=1}$ NV spin state projections (see SI). The two outermost peaks in the spectrum correspond to the hyperfine lines of P1 centers aligned with $B_0$ and the two strongest peaks correspond to the hyperfine lines of the three other P1 center orientations that are magnetically equivalent in the given magnetic field arrangement. The small central peaks can be explained by additional, weakly allowed transitions that appear between hyperfine states of the coupled electron spin and nuclear spin of the P1 center and are further explained in the supplemental information. We emphasize that with 145 keV electrons (dosage $\leq$ 10$^{22}$ e$^-$/cm$^2$) and with 200 keV electrons (dosage $\leq$ 10$^{21}$ e$^-$/cm$^2$) we do not see evidence of $g=2$ spins at 500 MHz nor other spins over a GHz frequency range, suggesting that other spin concentrations besides P1 centers are comparatively low.  It should also be noted that we would likely not see $g=2$ spins if their line width was broad as seen by Tetienne \textit{et al} \cite{tetienne2018spin}. 

We next show how we measure P1 density using the pulse sequence in Fig. \ref{fig:DEER_Hahn_echo}b. By fixing the frequency of the pump $\pi$-pulse to one of the aligned P1 spin transitions, $f_{\text{pump}}$ in Fig. \ref{fig:DEER_Hahn_echo}a, and sweeping the pump pulse in time with respect to the final $\pi/2$-pulse in the NV Hahn echo sequence, we can control the degree to which we recouple the P1 center spins. In such a measurement, the ensemble average of the static dipolar interactions will give rise to a mono-exponential echo decay of the coherence of the probe spins (SE) that depends on the density $n_{\text{bath}}$ of inverted bath spins\cite{salikhov1981theory,romanelli1997evaluation} and the time they were re-coupled into the Hahn echo, T:

\begin{align}\label{eq:DEER_decay}
SE(n,T) \propto \exp(- A \gamma_{\text{NV}} \gamma_{\text{bath}} \cdot  n_{\text{bath}} \cdot T),
\end{align}
where $\gamma$ denotes the spin species' gyromagnetic ratio and $A$ a numerical prefactor that depends on the angle between the quantization axes of the probe and pump spins (in case S $>$ 1). Both the P1 and the NV spin densities can be measured with this technique.

Because $B_0$ is large compared to the hyperfine splitting of the P1 centers and the NV probe is aligned to $B_0$, we assume the quantization axes of the NV center probe and P1 centers are the same. Therefore, the prefactor $A \gamma_{\text{NV}} \gamma_{\text{bath}}$ becomes 292 kHz/ppm\cite{stepanov2016determination}. To further improve the accuracy of our measurement we correct our estimate of $n_\text{bath}$ to account for the fidelity of our pump $\pi$ pulse, which we calculate considering the spatial inhomogeneity of the microwave magnetic field and the coherence time of the pump spins (see SI). We note that Eq. \ref{eq:DEER_decay} holds true when the correlation time of the spins ($\tau_c$) is longer than the measurement time ($2\tau < \tau_c$), known as the quasi-static regime; in this regime, the decay in Eq. \ref{eq:DEER_decay} is due to static dipolar interactions. The $\tau_c$ of the entire diamond spin bath was measured longer than a millisecond in all three samples (see SI).

In Fig. \ref{fig:DEER_Hahn_echo}b we plot the NV coherence decay versus the re-coupling time for two different areas on sample C041, one irradiated with electrons and one non-irradiated. Fits to the data using Eq. \ref{eq:DEER_decay} give $n_{\text{P1}}$ = 10 ppm for the electron-irradiated area, and $n_{\text{P1}}$ = 20 ppm for the non-irradiated area. The factor of 2 decrease in $n_{\text{P1}}$ results from irradiation-induced conversion of P1 centers. 


We also use the DEER technique to measure NV center spin density $n_\text{NV}$, where we use the NV center spins aligned with $B_0$ as pump spins and the other NV orientations as probe spins. In the analysis of the data, we include a correction to the prefactor $A$ due to the large NV center zero-field splitting that determines the NV quantization axis at the low $B_0$ used in these studies (see SI). In this paper we use both DEER and instantaneous diffusion, explained in the next section, to extract $n_\text{NV}$.


\subsection{\label{subsec:NVinteractions}NV density detection via instantaneous diffusion}

\begin{figure}
\includegraphics[width=\columnwidth]{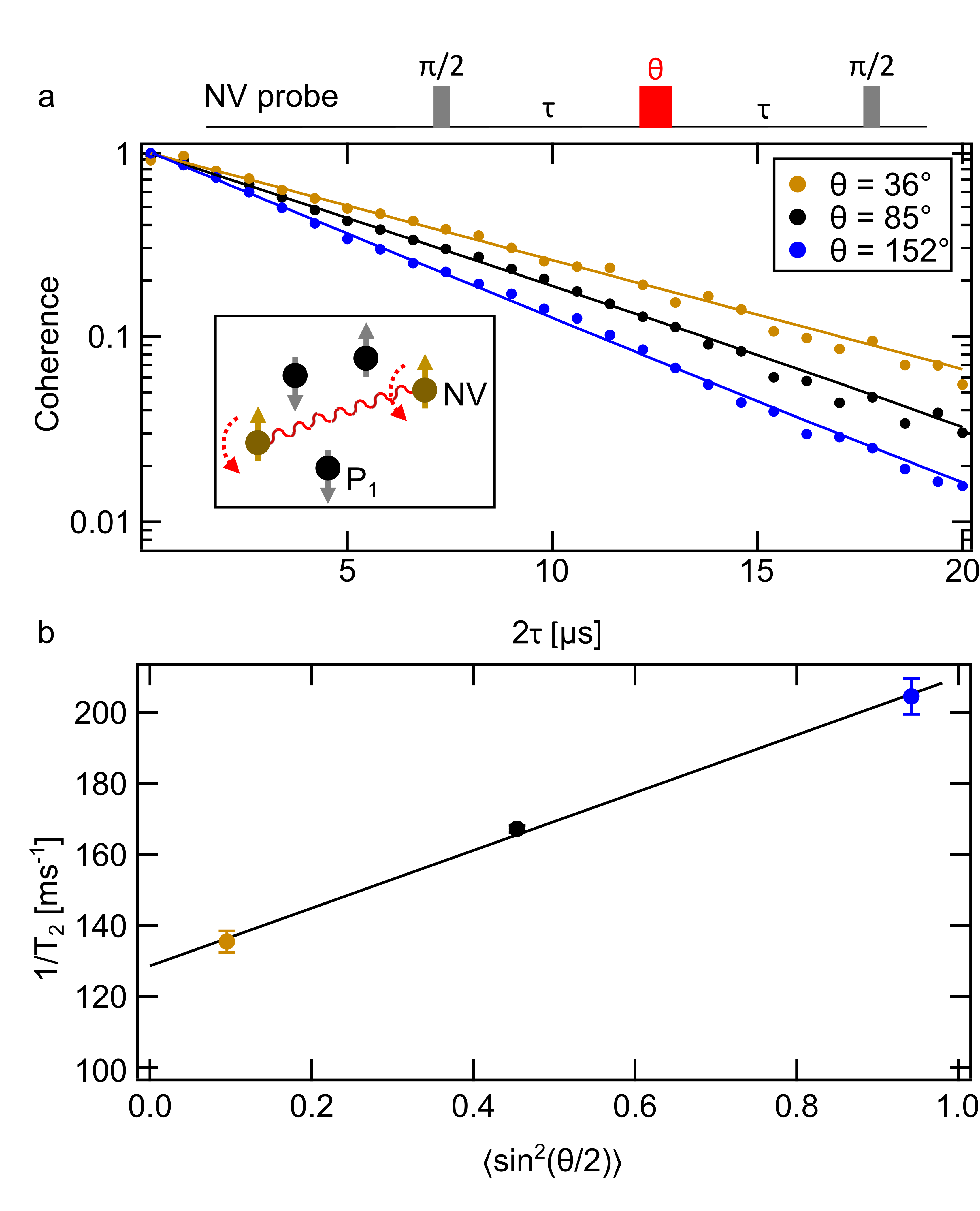}
\caption{Quantifying NV-NV interactions with instantaneous diffusion. (a) Hahn echo coherence measurements on NV spins taken with three different flip angles $\theta$ of the central pulse. As $\theta$ deviates from $\pi$ the coherence of the NV spin bath is increased, indicating that dipolar interactions between probed spins limit the coherence (so-called instantaneous diffusion). Solid lines are exponential fits to the data.
(b) Inverse coherence time, extracted from fits to the data in (a) plotted as a function of averaged inversion pulse fidelity $\langle \sin^2(\theta/2) \rangle$. Linear fit to the data (solid line) yields the density of probed NV spins. Data taken on sample C041, 200 keV, 10$^{20}$ e$^-$/cm$^2$.}
\label{fig:inst_diff}
\end{figure}

At high NV densities compared to all other spin defect densities, NV-NV spin interactions begin to dominate the NV spin decoherence. In this regime, instantaneous diffusion can be used to probe NV-NV spin interactions and hence $n_\text{NV}$. This powerful technique has been used to detect interactions in phosphorus spins in silicon \cite{tyryshkin2012electron}, P1 centers in diamond \cite{wilson2018multi,stepanov2016determination}, and NV centers in diamond \cite{ROSE201732}. In an instantaneous diffusion measurement (Fig. \ref{fig:inst_diff}), Hahn echo sequences are performed on NV center spins while the angle, and hence the fidelity, of the central inversion pulse is varied  \cite{1979Tdes}. The central pulse inhibits the decoupling of the probed spins' mutual dipolar interactions, resulting in decoherence in a process known as instantaneous diffusion. Reducing the angle of the central pulse reduces instantaneous diffusion, thus increasing the coherence of the ensemble.  It should be noted that as the fidelity of the central pulse decreases, the signal of the echo also decreases. For nonunity pulse fidelities, we use phase cycling (see SI) to remove the effects of free induction decay. 

In analogy to eq. \ref{eq:DEER_decay}, the Hahn echo signal is described by 
\begin{align}\label{eq:ID_decay}
SE(n_\text{NV}^p,\theta,\tau) \propto \exp \left(-A\gamma^2_{\text{NV}} \cdot  n_\text{NV}^p \cdot \langle \sin^2 (\theta/2) \rangle \cdot \tau \right),
\end{align}
where 
$n_\text{NV}^p$ is the density of the NV spin class being probed ($n_\text{NV}^p=n_\text{NV}/4$),  and $\theta$ is the central pulse's flip angle \cite{klauder1962spectral,salikhov1981theory}. For the monoexponential coherence decays in Fig. \ref{fig:inst_diff}a taken on sample C041, Eq. \ref{eq:ID_decay} states that 1/$T_2 \propto \langle \sin^2 (\theta/2) \rangle$ and hence the slope of the linear fit to the data in Fig. 4b yields the interaction strength of the probed NV center spin class to be $(80 \pm\,3)$ kHz, corresponding to a total NV density $n_\text{NV}=(2.2 \pm 0.2)$ ppm, which takes into account a factor of 4 to include all 4 NV classes.

Measuring spin densities with instantaneous diffusion is useful in situations where only a single MW frequency can be applied \cite{Milov1998}, but can be challenging to interpret in the presence of electron spin echo envelope modulation (ESEEM) effects from, \textit{e.g.}, a $^{13}$C nuclear spin \cite{rowan1965electron}. The measurements shown in Fig. \ref{fig:inst_diff} are done on a $^{12}$C-isotopically pure diamond sample that shows a mono-exponential Hahn echo decay.  
The DEER technique is preferable in the presence of ESEEM because the probe delay time, $\tau$, remains fixed during the sequence. DEER techniques are also more sensitive at lower NV spin densities.



\begin{figure*}[ht]
\includegraphics[scale=1.0]{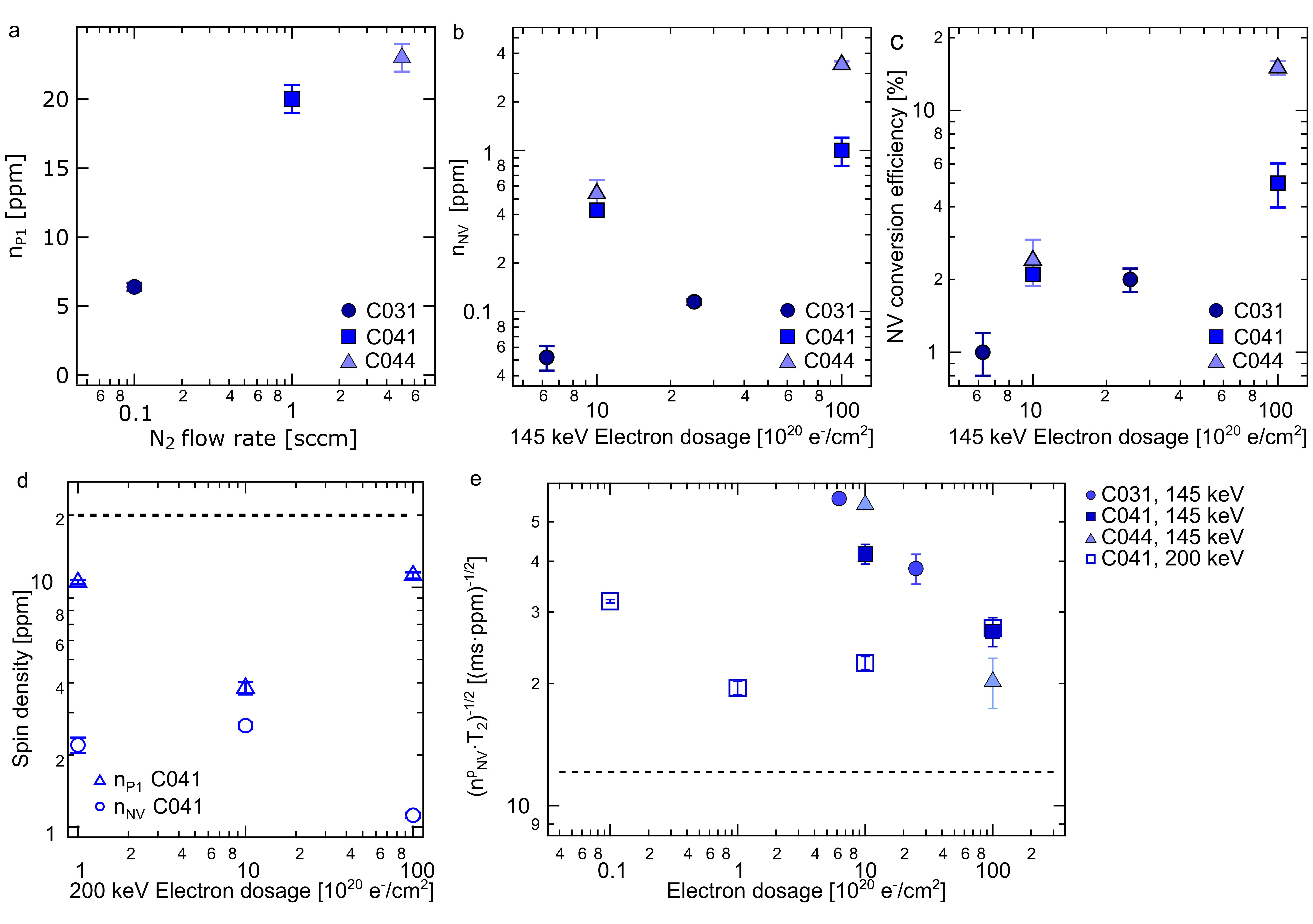}
\caption{Summary of spin density and optimization of magnetic sensitivity. (a) P1 spin density as a function of nitrogen gas flow during growth for the three samples studied. Data is taken in lightly irradiated regions, where $n_\text{P1}$ is minimally altered by irradiation.  (b) NV density and (c) NV conversion efficiency as a function of electron irradiation dose for 145 keV electrons. (d) P1 and NV spin densities as a function of electron irradiation dose for 200 keV electrons on sample C041. (e) Product of the probed NV spin density ($n_\text{NV}^p = n_\text{NV}/4$) and coherence time T$_2$, which relates to the magnetic sensitivity. The dashed line indicates the limit achieved when decoherence is solely due to static dipolar interaction between probed NV centers. We note that there are two almost-overlapping data points at $10^{22}$ e$^-$/cm$^2$ for C041, 145keV and C041, 200 keV.}
\label{fig:Fig5}
\end{figure*}



\section{\label{sec:discussion}Discussion}
\subsection{\label{subsec:disc_P1_limit}Optimizing NV ensemble magnetic sensitivity}

We use the techniques described in section \ref{sec:measurements} to explore a wide phase space of NV center formation parameters to optimize the magnetic sensitivity of our NV ensembles. We summarize our results in Fig. \ref{fig:Fig5} a-d, which demonstrate our control over $n_\text{P1}$ and NV center density and conversion efficiency, and showing our ability to form an NV ensemble optimized for magnetic sensing. 


Figure \ref{fig:Fig5}a plots $n_\text{P1}$ in three diamond films grown with different N$_2$ flow rates during CVD growth. Spin density measurements, using Hahn echo-based DEER described in section \ref{subsec:DEER-sequence}, were performed on lightly irradiated $(< 10^{18}$ e$^-$/cm$^2$) areas of the samples where $n_\text{P1}$ is minimally affected by irradiation. Increasing the N$_2$ flow rate from 0.1 sccm to 5 sccm increases $n_\text{P1}$ in the grown diamond from 6 ppm to 22 ppm. Because the change in $n_\text{P1}$ from 1 sccm to 5 sccm  N$_2$ is smaller than from 0.1 sccm to 1 sccm N$_2$, we may be seeing some saturation in the incorporation of substitutional nitrogen. 

Introducing vacancies into the diamond via electron irradiation from a TEM allows for precise control over the electron dose and energy. We irradiate using 145 keV electrons, just at the threshold energy for vacancy creation in diamond \cite{campbell2000radiation,mclellan2016,koike1992displacement,clark1961optical}, and 200 keV electrons. 
The 145 keV (200 keV) electrons form vacancies in the first micrometer (60 $\mu$m) of the diamond \cite{mclellan2016}. 

In Fig. \ref{fig:Fig5}b-c we show control over both $n_\text{NV}$ (Fig.  \ref{fig:Fig5}b) and conversion of P1 centers to NV centers (Fig.  \ref{fig:Fig5}c) by tuning the electron irradiation dosage from $6.3 \cdot 10^{20}$ to $10^{22}$ e$^-$/cm$^{2}$ at 145 keV. The NV conversion efficiency is defined as the ratio of $n_\text{NV}$ to the starting $n_\text{P1}$. Both $n_\text{NV}$ and NV conversion efficiency increase with electron dosage without apparent saturation. 
Higher electron dosages were not explored because of the prohibitively long electron irradiation times needed.  
NV densities were measured using Hahn echo-based sensing described in section \ref{subsec:DEER-sequence} and \ref{subsec:NVinteractions}.

Figure \ref{fig:Fig5}d, which plots $n_\text{P1}$ and $n_\text{NV}$ as a function of electron dosage $(10^{20}-10^{22}$ e$^-$/cm$^{2}$) for 200 keV electrons, shows markedly different behavior. Firstly, the NV conversion efficiency is significantly larger for 200 keV than 145 keV electrons for dosages at and below $10^{21}$ e$^-$/cm$^{2}$. For example, on C041, we achieve only  2$\%$ NV conversion efficiency with $10^{21}$ e$^-$/cm$^{2}$, 145 keV electrons, whereas we achieve 13 $\%$ NV conversion efficiency with 200 keV electrons for the same dose. The dependence of $n_\text{NV}$ on electron dosage is also different: $n_\text{NV}$ initially increases with dosage and then drops at $10^{22}$ e$^-$/cm$^{2}$. 

Importantly, Fig. \ref{fig:Fig5}e shows how we are able to hone in on an optimized set of growth and irradiation parameters that realize an NV ensemble whose magnetic sensitivity is nearly limited by dipolar interactions between the sensor spins themselves. In this NV-NV interaction-limited regime, the parameter $(n_{\text{NV}}^p\cdot T_2)^{-1/2}$, which determines the ensemble sensitivity, reaches a theoretical limit of 12.1 $(\text{ms}\cdot \text{ppm})^{-1/2}$ \cite{wang2013spin,romanelli1997evaluation} because $T_2 \sim 1/n_\text{NV}^{p}$ when dominated by NV-NV dipolar interactions. The dashed horizontal line in Fig. \ref{fig:Fig5}e indicates this theoretical value. We note that we use $n_\text{NV}^p$ here to indicate the density of one NV spin class, which acts as the sensor spins. The NV spins of other orientations neither contribute to sensing, nor to decoherence, because they are all assumed polarized into the nonmagnetic $\ket{m_s=0}$ state.

The data in Fig. \ref{fig:Fig5}e plots $(n_{\text{NV}}^p\cdot T_2)^{-1/2}$ as a function of electron irradiation dosage for 145 keV and 200 keV electrons. For 145 keV electrons (filled data points) the sensitivity of our samples improves with increased electron dosage. This improvement occurs because $n_\text{NV}$ increases with dosage while $T_2$ does not change appreciably ($T_2$ results are presented in the SI). In contrast, for 200 keV irradiation (open data points), the sensitivity improves with dosage up to 10$^{20}$ e$^-$/cm$^2$, and then starts to degrade. This sensitivity degradation is due to a reduction in both $n_\text{NV}$, as seen in Fig. \ref{fig:Fig5}d, and in $T_2$, which we explain by the proposed presence of di-vacancy or other vacancy-related defects that limit the NV$^-$ coherence.

A key result is that with a dose of 10$^{20}$ e$^-$/cm$^2$, 200 keV electrons (sample C041) we produce an ensemble with $n_\text{NV}$  = 2.2 ppm, $n_\text{NV}^p$ = 0.55 ppm,  T$_2$ = 4.9 $\mu$s, and  $(n_{\text{NV}}\cdot T_2)^{-1/2}$ = (19 $\pm 0.4$) $(\text{ms}\cdot \text{ppm})^{-1/2}$, which is a factor of 1.6 away from the optimum dipolar-limited sensitivity. Assuming a typical photon count rate of 150 kCnt/($\text{s}\cdot\text{NV}$) and a fluorescence contrast of 0.85 between the $\ket{m_s=0}$ and $\ket{m_s=1}$ states, we estimate a sensitivity of 370 pT/$\sqrt{\text{Hz}}$ in a 1 $\mu m^3$ sensing volume \cite{pham2013magnetic}.

\subsection{\label{subsec:conclusions}Conclusions and future work}

Producing a highly coherent, depth-confined ensemble of NV centers is critical for applications in wide-field magnetometry. Having reached nearly dipolar-interaction-limited coherence times, our NV ensembles are excellent platforms to augment magnetometry experiments with advanced sensing techniques such as multiplexing for vector magnetometry \cite{schloss2018simultaneous} and using quantum correlations \cite{cappellaro2009quantum,choi2017quantum,Zhou2018} 
to go beyond the dipolar-interaction limit to sensitivity.

Pathways towards forming NV ensembles that eke out a further 1.6x improvement in sensitivity to reach the theoretical dipolar-interaction limit of $(n_{\text{NV}}^p\cdot T_2)^{-1/2} = 12.1 (\text{ms}\cdot \text{ppm})^{-1/2}$ could include post-annealing above 1100 $^\circ$C to remove divacancies or other defects \cite{naydenov2010increasing,chu2014coherent}, or irradiating while annealing to further improve the NV conversion efficiency \cite{choi2017depolarization}. 

Other potential applications that could utilize these optimized high-density NV ensembles include the exploration of driven, strongly interacting spin systems \cite{kucsko2018critical,yao2014many,choi2017observation,rovny2018observation} and hybrid quantum systems that couple NV spin ensembles to superconducting qubits \cite{putz2014protecting}. 




\bibliography{Bib}

\begin{thebibliography}{42}%
\makeatletter
\providecommand \@ifxundefined [1]{%
 \@ifx{#1\undefined}
}%
\providecommand \@ifnum [1]{%
 \ifnum #1\expandafter \@firstoftwo
 \else \expandafter \@secondoftwo
 \fi
}%
\providecommand \@ifx [1]{%
 \ifx #1\expandafter \@firstoftwo
 \else \expandafter \@secondoftwo
 \fi
}%
\providecommand \natexlab [1]{#1}%
\providecommand \enquote  [1]{``#1''}%
\providecommand \bibnamefont  [1]{#1}%
\providecommand \bibfnamefont [1]{#1}%
\providecommand \citenamefont [1]{#1}%
\providecommand \href@noop [0]{\@secondoftwo}%
\providecommand \href [0]{\begingroup \@sanitize@url \@href}%
\providecommand \@href[1]{\@@startlink{#1}\@@href}%
\providecommand \@@href[1]{\endgroup#1\@@endlink}%
\providecommand \@sanitize@url [0]{\catcode `\\12\catcode `\$12\catcode
  `\&12\catcode `\#12\catcode `\^12\catcode `\_12\catcode `\%12\relax}%
\providecommand \@@startlink[1]{}%
\providecommand \@@endlink[0]{}%
\providecommand \url  [0]{\begingroup\@sanitize@url \@url }%
\providecommand \@url [1]{\endgroup\@href {#1}{\urlprefix }}%
\providecommand \urlprefix  [0]{URL }%
\providecommand \Eprint [0]{\href }%
\providecommand \doibase [0]{http://dx.doi.org/}%
\providecommand \selectlanguage [0]{\@gobble}%
\providecommand \bibinfo  [0]{\@secondoftwo}%
\providecommand \bibfield  [0]{\@secondoftwo}%
\providecommand \translation [1]{[#1]}%
\providecommand \BibitemOpen [0]{}%
\providecommand \bibitemStop [0]{}%
\providecommand \bibitemNoStop [0]{.\EOS\space}%
\providecommand \EOS [0]{\spacefactor3000\relax}%
\providecommand \BibitemShut  [1]{\csname bibitem#1\endcsname}%
\let\auto@bib@innerbib\@empty
\bibitem [{\citenamefont {Doherty}\ \emph {et~al.}(2013)\citenamefont
  {Doherty}, \citenamefont {Manson}, \citenamefont {Delaney}, \citenamefont
  {Jelezko}, \citenamefont {Wrachtrup},\ and\ \citenamefont
  {Hollenberg}}]{doherty2013nitrogen}%
  \BibitemOpen
  \bibfield  {author} {\bibinfo {author} {\bibfnamefont {M.~W.}\ \bibnamefont
  {Doherty}}, \bibinfo {author} {\bibfnamefont {N.~B.}\ \bibnamefont {Manson}},
  \bibinfo {author} {\bibfnamefont {P.}~\bibnamefont {Delaney}}, \bibinfo
  {author} {\bibfnamefont {F.}~\bibnamefont {Jelezko}}, \bibinfo {author}
  {\bibfnamefont {J.}~\bibnamefont {Wrachtrup}}, \ and\ \bibinfo {author}
  {\bibfnamefont {L.~C.}\ \bibnamefont {Hollenberg}},\ }\href@noop {}
  {\bibfield  {journal} {\bibinfo  {journal} {Physics Reports}\ }\textbf
  {\bibinfo {volume} {528}},\ \bibinfo {pages} {1} (\bibinfo {year}
  {2013})}\BibitemShut {NoStop}%
\bibitem [{\citenamefont {Pelliccione}\ \emph {et~al.}(2016)\citenamefont
  {Pelliccione}, \citenamefont {Jenkins}, \citenamefont {Ovartchaiyapong},
  \citenamefont {Reetz}, \citenamefont {Emmanouilidou}, \citenamefont {Ni},\
  and\ \citenamefont {Jayich}}]{pelliccione2016scanned}%
  \BibitemOpen
  \bibfield  {author} {\bibinfo {author} {\bibfnamefont {M.}~\bibnamefont
  {Pelliccione}}, \bibinfo {author} {\bibfnamefont {A.}~\bibnamefont
  {Jenkins}}, \bibinfo {author} {\bibfnamefont {P.}~\bibnamefont
  {Ovartchaiyapong}}, \bibinfo {author} {\bibfnamefont {C.}~\bibnamefont
  {Reetz}}, \bibinfo {author} {\bibfnamefont {E.}~\bibnamefont
  {Emmanouilidou}}, \bibinfo {author} {\bibfnamefont {N.}~\bibnamefont {Ni}}, \
  and\ \bibinfo {author} {\bibfnamefont {A.~C.~B.}\ \bibnamefont {Jayich}},\
  }\href@noop {} {\bibfield  {journal} {\bibinfo  {journal} {Nature
  Nanotechnology}\ }\textbf {\bibinfo {volume} {11}},\ \bibinfo {pages} {700}
  (\bibinfo {year} {2016})}\BibitemShut {NoStop}%
\bibitem [{\citenamefont {Maletinsky}\ \emph {et~al.}(2012)\citenamefont
  {Maletinsky}, \citenamefont {Hong}, \citenamefont {Grinolds}, \citenamefont
  {Hausmann}, \citenamefont {Lukin}, \citenamefont {Walsworth}, \citenamefont
  {Loncar},\ and\ \citenamefont {Yacoby}}]{maletinsky2012robust}%
  \BibitemOpen
  \bibfield  {author} {\bibinfo {author} {\bibfnamefont {P.}~\bibnamefont
  {Maletinsky}}, \bibinfo {author} {\bibfnamefont {S.}~\bibnamefont {Hong}},
  \bibinfo {author} {\bibfnamefont {M.~S.}\ \bibnamefont {Grinolds}}, \bibinfo
  {author} {\bibfnamefont {B.}~\bibnamefont {Hausmann}}, \bibinfo {author}
  {\bibfnamefont {M.~D.}\ \bibnamefont {Lukin}}, \bibinfo {author}
  {\bibfnamefont {R.~L.}\ \bibnamefont {Walsworth}}, \bibinfo {author}
  {\bibfnamefont {M.}~\bibnamefont {Loncar}}, \ and\ \bibinfo {author}
  {\bibfnamefont {A.}~\bibnamefont {Yacoby}},\ }\href@noop {} {\bibfield
  {journal} {\bibinfo  {journal} {Nature nanotechnology}\ }\textbf {\bibinfo
  {volume} {7}},\ \bibinfo {pages} {320} (\bibinfo {year} {2012})}\BibitemShut
  {NoStop}%
\bibitem [{\citenamefont {Thiel}\ \emph {et~al.}(2016)\citenamefont {Thiel},
  \citenamefont {Rohner}, \citenamefont {Ganzhorn}, \citenamefont {Appel},
  \citenamefont {Neu}, \citenamefont {M{\"u}ller}, \citenamefont {Kleiner},
  \citenamefont {Koelle},\ and\ \citenamefont
  {Maletinsky}}]{thiel2016quantitative}%
  \BibitemOpen
  \bibfield  {author} {\bibinfo {author} {\bibfnamefont {L.}~\bibnamefont
  {Thiel}}, \bibinfo {author} {\bibfnamefont {D.}~\bibnamefont {Rohner}},
  \bibinfo {author} {\bibfnamefont {M.}~\bibnamefont {Ganzhorn}}, \bibinfo
  {author} {\bibfnamefont {P.}~\bibnamefont {Appel}}, \bibinfo {author}
  {\bibfnamefont {E.}~\bibnamefont {Neu}}, \bibinfo {author} {\bibfnamefont
  {B.}~\bibnamefont {M{\"u}ller}}, \bibinfo {author} {\bibfnamefont
  {R.}~\bibnamefont {Kleiner}}, \bibinfo {author} {\bibfnamefont
  {D.}~\bibnamefont {Koelle}}, \ and\ \bibinfo {author} {\bibfnamefont
  {P.}~\bibnamefont {Maletinsky}},\ }\href@noop {} {\bibfield  {journal}
  {\bibinfo  {journal} {Nature Nanotechnology}\ }\textbf {\bibinfo {volume}
  {11}},\ \bibinfo {pages} {677} (\bibinfo {year} {2016})}\BibitemShut
  {NoStop}%
\bibitem [{\citenamefont {Barry}\ \emph {et~al.}(2016)\citenamefont {Barry},
  \citenamefont {Turner}, \citenamefont {Schloss}, \citenamefont {Glenn},
  \citenamefont {Song}, \citenamefont {Lukin}, \citenamefont {Park},\ and\
  \citenamefont {Walsworth}}]{barry2016optical}%
  \BibitemOpen
  \bibfield  {author} {\bibinfo {author} {\bibfnamefont {J.~F.}\ \bibnamefont
  {Barry}}, \bibinfo {author} {\bibfnamefont {M.~J.}\ \bibnamefont {Turner}},
  \bibinfo {author} {\bibfnamefont {J.~M.}\ \bibnamefont {Schloss}}, \bibinfo
  {author} {\bibfnamefont {D.~R.}\ \bibnamefont {Glenn}}, \bibinfo {author}
  {\bibfnamefont {Y.}~\bibnamefont {Song}}, \bibinfo {author} {\bibfnamefont
  {M.~D.}\ \bibnamefont {Lukin}}, \bibinfo {author} {\bibfnamefont
  {H.}~\bibnamefont {Park}}, \ and\ \bibinfo {author} {\bibfnamefont {R.~L.}\
  \bibnamefont {Walsworth}},\ }\href@noop {} {\bibfield  {journal} {\bibinfo
  {journal} {Proceedings of the National Academy of Sciences}\ }\textbf
  {\bibinfo {volume} {113}},\ \bibinfo {pages} {14133} (\bibinfo {year}
  {2016})}\BibitemShut {NoStop}%
\bibitem [{\citenamefont {Tetienne}\ \emph {et~al.}(2018)\citenamefont
  {Tetienne}, \citenamefont {de~Gille}, \citenamefont {Broadway}, \citenamefont
  {Teraji}, \citenamefont {Lillie}, \citenamefont {McCoey}, \citenamefont
  {Dontschuk}, \citenamefont {Hall}, \citenamefont {Stacey}, \citenamefont
  {Simpson},\ and\ \citenamefont {Hollenberg}}]{tetienne2018spin}%
  \BibitemOpen
  \bibfield  {author} {\bibinfo {author} {\bibfnamefont {J.-P.}\ \bibnamefont
  {Tetienne}}, \bibinfo {author} {\bibfnamefont {R.~W.}\ \bibnamefont
  {de~Gille}}, \bibinfo {author} {\bibfnamefont {D.~A.}\ \bibnamefont
  {Broadway}}, \bibinfo {author} {\bibfnamefont {T.}~\bibnamefont {Teraji}},
  \bibinfo {author} {\bibfnamefont {S.~E.}\ \bibnamefont {Lillie}}, \bibinfo
  {author} {\bibfnamefont {J.~M.}\ \bibnamefont {McCoey}}, \bibinfo {author}
  {\bibfnamefont {N.}~\bibnamefont {Dontschuk}}, \bibinfo {author}
  {\bibfnamefont {L.~T.}\ \bibnamefont {Hall}}, \bibinfo {author}
  {\bibfnamefont {A.}~\bibnamefont {Stacey}}, \bibinfo {author} {\bibfnamefont
  {D.~A.}\ \bibnamefont {Simpson}}, \ and\ \bibinfo {author} {\bibfnamefont
  {L.~C.~L.}\ \bibnamefont {Hollenberg}},\ }\href {\doibase
  10.1103/PhysRevB.97.085402} {\bibfield  {journal} {\bibinfo  {journal}
  {Physical Review B}\ }\textbf {\bibinfo {volume} {97}},\ \bibinfo {pages}
  {085402} (\bibinfo {year} {2018})}\BibitemShut {NoStop}%
\bibitem [{\citenamefont {Simpson}\ \emph {et~al.}(2017)\citenamefont
  {Simpson}, \citenamefont {Ryan}, \citenamefont {Hall}, \citenamefont
  {Panchenko}, \citenamefont {Drew}, \citenamefont {Petrou}, \citenamefont
  {Donnelly}, \citenamefont {Mulvaney},\ and\ \citenamefont
  {Hollenberg}}]{simpson2017electron}%
  \BibitemOpen
  \bibfield  {author} {\bibinfo {author} {\bibfnamefont {D.~A.}\ \bibnamefont
  {Simpson}}, \bibinfo {author} {\bibfnamefont {R.~G.}\ \bibnamefont {Ryan}},
  \bibinfo {author} {\bibfnamefont {L.~T.}\ \bibnamefont {Hall}}, \bibinfo
  {author} {\bibfnamefont {E.}~\bibnamefont {Panchenko}}, \bibinfo {author}
  {\bibfnamefont {S.~C.}\ \bibnamefont {Drew}}, \bibinfo {author}
  {\bibfnamefont {S.}~\bibnamefont {Petrou}}, \bibinfo {author} {\bibfnamefont
  {P.~S.}\ \bibnamefont {Donnelly}}, \bibinfo {author} {\bibfnamefont
  {P.}~\bibnamefont {Mulvaney}}, \ and\ \bibinfo {author} {\bibfnamefont
  {L.~C.}\ \bibnamefont {Hollenberg}},\ }\href@noop {} {\bibfield  {journal}
  {\bibinfo  {journal} {Nature Communications}\ }\textbf {\bibinfo {volume}
  {8}},\ \bibinfo {pages} {458} (\bibinfo {year} {2017})}\BibitemShut {NoStop}%
\bibitem [{\citenamefont {Wolf}\ \emph {et~al.}(2015)\citenamefont {Wolf},
  \citenamefont {Neumann}, \citenamefont {Nakamura}, \citenamefont {Sumiya},
  \citenamefont {Ohshima}, \citenamefont {Isoya},\ and\ \citenamefont
  {Wrachtrup}}]{wolf2015subpicotesla}%
  \BibitemOpen
  \bibfield  {author} {\bibinfo {author} {\bibfnamefont {T.}~\bibnamefont
  {Wolf}}, \bibinfo {author} {\bibfnamefont {P.}~\bibnamefont {Neumann}},
  \bibinfo {author} {\bibfnamefont {K.}~\bibnamefont {Nakamura}}, \bibinfo
  {author} {\bibfnamefont {H.}~\bibnamefont {Sumiya}}, \bibinfo {author}
  {\bibfnamefont {T.}~\bibnamefont {Ohshima}}, \bibinfo {author} {\bibfnamefont
  {J.}~\bibnamefont {Isoya}}, \ and\ \bibinfo {author} {\bibfnamefont
  {J.}~\bibnamefont {Wrachtrup}},\ }\href@noop {} {\bibfield  {journal}
  {\bibinfo  {journal} {Physical Review X}\ }\textbf {\bibinfo {volume} {5}},\
  \bibinfo {pages} {041001} (\bibinfo {year} {2015})}\BibitemShut {NoStop}%
\bibitem [{\citenamefont {Acosta}\ \emph {et~al.}(2009)\citenamefont {Acosta},
  \citenamefont {Bauch}, \citenamefont {Ledbetter}, \citenamefont {Santori},
  \citenamefont {Fu}, \citenamefont {Barclay}, \citenamefont {Beausoleil},
  \citenamefont {Linget}, \citenamefont {Roch}, \citenamefont {Treussart} \emph
  {et~al.}}]{acosta2009diamonds}%
  \BibitemOpen
  \bibfield  {author} {\bibinfo {author} {\bibfnamefont {V.}~\bibnamefont
  {Acosta}}, \bibinfo {author} {\bibfnamefont {E.}~\bibnamefont {Bauch}},
  \bibinfo {author} {\bibfnamefont {M.}~\bibnamefont {Ledbetter}}, \bibinfo
  {author} {\bibfnamefont {C.}~\bibnamefont {Santori}}, \bibinfo {author}
  {\bibfnamefont {K.-M.}\ \bibnamefont {Fu}}, \bibinfo {author} {\bibfnamefont
  {P.}~\bibnamefont {Barclay}}, \bibinfo {author} {\bibfnamefont
  {R.}~\bibnamefont {Beausoleil}}, \bibinfo {author} {\bibfnamefont
  {H.}~\bibnamefont {Linget}}, \bibinfo {author} {\bibfnamefont
  {J.}~\bibnamefont {Roch}}, \bibinfo {author} {\bibfnamefont {F.}~\bibnamefont
  {Treussart}},  \emph {et~al.},\ }\href@noop {} {\bibfield  {journal}
  {\bibinfo  {journal} {Physical Review B}\ }\textbf {\bibinfo {volume} {80}},\
  \bibinfo {pages} {115202} (\bibinfo {year} {2009})}\BibitemShut {NoStop}%
\bibitem [{\citenamefont {Bauch}\ \emph {et~al.}(2018)\citenamefont {Bauch},
  \citenamefont {Hart}, \citenamefont {Schloss}, \citenamefont {Turner},
  \citenamefont {Barry}, \citenamefont {Kehayias}, \citenamefont {Singh},\ and\
  \citenamefont {Walsworth}}]{bauch2018ultralong}%
  \BibitemOpen
  \bibfield  {author} {\bibinfo {author} {\bibfnamefont {E.}~\bibnamefont
  {Bauch}}, \bibinfo {author} {\bibfnamefont {C.~A.}\ \bibnamefont {Hart}},
  \bibinfo {author} {\bibfnamefont {J.~M.}\ \bibnamefont {Schloss}}, \bibinfo
  {author} {\bibfnamefont {M.~J.}\ \bibnamefont {Turner}}, \bibinfo {author}
  {\bibfnamefont {J.~F.}\ \bibnamefont {Barry}}, \bibinfo {author}
  {\bibfnamefont {P.}~\bibnamefont {Kehayias}}, \bibinfo {author}
  {\bibfnamefont {S.}~\bibnamefont {Singh}}, \ and\ \bibinfo {author}
  {\bibfnamefont {R.~L.}\ \bibnamefont {Walsworth}},\ }\href {\doibase
  10.1103/PhysRevX.8.031025} {\bibfield  {journal} {\bibinfo  {journal}
  {Physical Review X}\ }\textbf {\bibinfo {volume} {8}},\ \bibinfo {pages}
  {031025} (\bibinfo {year} {2018})}\BibitemShut {NoStop}%
\bibitem [{\citenamefont {Cappellaro}\ and\ \citenamefont
  {Lukin}(2009)}]{cappellaro2009quantum}%
  \BibitemOpen
  \bibfield  {author} {\bibinfo {author} {\bibfnamefont {P.}~\bibnamefont
  {Cappellaro}}\ and\ \bibinfo {author} {\bibfnamefont {M.~D.}\ \bibnamefont
  {Lukin}},\ }\href@noop {} {\bibfield  {journal} {\bibinfo  {journal}
  {Physical Review A}\ }\textbf {\bibinfo {volume} {80}},\ \bibinfo {pages}
  {032311} (\bibinfo {year} {2009})}\BibitemShut {NoStop}%
\bibitem [{\citenamefont {Choi}\ \emph
  {et~al.}(2017{\natexlab{a}})\citenamefont {Choi}, \citenamefont {Yao},\ and\
  \citenamefont {Lukin}}]{choi2017quantum}%
  \BibitemOpen
  \bibfield  {author} {\bibinfo {author} {\bibfnamefont {S.}~\bibnamefont
  {Choi}}, \bibinfo {author} {\bibfnamefont {N.~Y.}\ \bibnamefont {Yao}}, \
  and\ \bibinfo {author} {\bibfnamefont {M.~D.}\ \bibnamefont {Lukin}},\
  }\href@noop {} {\bibfield  {journal} {\bibinfo  {journal} {arXiv preprint
  arXiv:1801.00042}\ } (\bibinfo {year} {2017}{\natexlab{a}})}\BibitemShut
  {NoStop}%
\bibitem [{\citenamefont {Zhou}\ \emph {et~al.}(2018)\citenamefont {Zhou},
  \citenamefont {Zhang}, \citenamefont {Preskill},\ and\ \citenamefont
  {Jiang}}]{Zhou2018}%
  \BibitemOpen
  \bibfield  {author} {\bibinfo {author} {\bibfnamefont {S.}~\bibnamefont
  {Zhou}}, \bibinfo {author} {\bibfnamefont {M.}~\bibnamefont {Zhang}},
  \bibinfo {author} {\bibfnamefont {J.}~\bibnamefont {Preskill}}, \ and\
  \bibinfo {author} {\bibfnamefont {L.}~\bibnamefont {Jiang}},\ }\href@noop {}
  {\bibfield  {journal} {\bibinfo  {journal} {Nature Communications}\ }\textbf
  {\bibinfo {volume} {9}},\ \bibinfo {pages} {78} (\bibinfo {year}
  {2018})}\BibitemShut {NoStop}%
\bibitem [{\citenamefont {Palyanov}\ \emph {et~al.}(2010)\citenamefont
  {Palyanov}, \citenamefont {Borzdov}, \citenamefont {Khokhryakov},
  \citenamefont {Kupriyanov},\ and\ \citenamefont
  {Sokol}}]{palyanov2010effect}%
  \BibitemOpen
  \bibfield  {author} {\bibinfo {author} {\bibfnamefont {Y.~N.}\ \bibnamefont
  {Palyanov}}, \bibinfo {author} {\bibfnamefont {Y.~M.}\ \bibnamefont
  {Borzdov}}, \bibinfo {author} {\bibfnamefont {A.~F.}\ \bibnamefont
  {Khokhryakov}}, \bibinfo {author} {\bibfnamefont {I.~N.}\ \bibnamefont
  {Kupriyanov}}, \ and\ \bibinfo {author} {\bibfnamefont {A.~G.}\ \bibnamefont
  {Sokol}},\ }\href@noop {} {\bibfield  {journal} {\bibinfo  {journal} {Crystal
  Growth \& Design}\ }\textbf {\bibinfo {volume} {10}},\ \bibinfo {pages}
  {3169} (\bibinfo {year} {2010})}\BibitemShut {NoStop}%
\bibitem [{\citenamefont {Naydenov}\ \emph {et~al.}(2010)\citenamefont
  {Naydenov}, \citenamefont {Reinhard}, \citenamefont {L{\"a}mmle},
  \citenamefont {Richter}, \citenamefont {Kalish}, \citenamefont
  {D’Haenens-Johansson}, \citenamefont {Newton}, \citenamefont {Jelezko},\
  and\ \citenamefont {Wrachtrup}}]{naydenov2010increasing}%
  \BibitemOpen
  \bibfield  {author} {\bibinfo {author} {\bibfnamefont {B.}~\bibnamefont
  {Naydenov}}, \bibinfo {author} {\bibfnamefont {F.}~\bibnamefont {Reinhard}},
  \bibinfo {author} {\bibfnamefont {A.}~\bibnamefont {L{\"a}mmle}}, \bibinfo
  {author} {\bibfnamefont {V.}~\bibnamefont {Richter}}, \bibinfo {author}
  {\bibfnamefont {R.}~\bibnamefont {Kalish}}, \bibinfo {author} {\bibfnamefont
  {U.~F.}\ \bibnamefont {D’Haenens-Johansson}}, \bibinfo {author}
  {\bibfnamefont {M.}~\bibnamefont {Newton}}, \bibinfo {author} {\bibfnamefont
  {F.}~\bibnamefont {Jelezko}}, \ and\ \bibinfo {author} {\bibfnamefont
  {J.}~\bibnamefont {Wrachtrup}},\ }\href@noop {} {\bibfield  {journal}
  {\bibinfo  {journal} {Applied Physics Letters}\ }\textbf {\bibinfo {volume}
  {97}},\ \bibinfo {pages} {242511} (\bibinfo {year} {2010})}\BibitemShut
  {NoStop}%
\bibitem [{\citenamefont {Ohno}\ \emph {et~al.}(2012)\citenamefont {Ohno},
  \citenamefont {Joseph~Heremans}, \citenamefont {Bassett}, \citenamefont
  {Myers}, \citenamefont {Toyli}, \citenamefont {Bleszynski~Jayich},
  \citenamefont {Palmstr{\o}m},\ and\ \citenamefont {Awschalom}}]{ohno2012}%
  \BibitemOpen
  \bibfield  {author} {\bibinfo {author} {\bibfnamefont {K.}~\bibnamefont
  {Ohno}}, \bibinfo {author} {\bibfnamefont {F.}~\bibnamefont
  {Joseph~Heremans}}, \bibinfo {author} {\bibfnamefont {L.~C.}\ \bibnamefont
  {Bassett}}, \bibinfo {author} {\bibfnamefont {B.~A.}\ \bibnamefont {Myers}},
  \bibinfo {author} {\bibfnamefont {D.~M.}\ \bibnamefont {Toyli}}, \bibinfo
  {author} {\bibfnamefont {A.~C.}\ \bibnamefont {Bleszynski~Jayich}}, \bibinfo
  {author} {\bibfnamefont {C.~J.}\ \bibnamefont {Palmstr{\o}m}}, \ and\
  \bibinfo {author} {\bibfnamefont {D.~D.}\ \bibnamefont {Awschalom}},\
  }\href@noop {} {\bibfield  {journal} {\bibinfo  {journal} {Applied Physics
  Letters}\ }\textbf {\bibinfo {volume} {101}},\ \bibinfo {pages} {082413}
  (\bibinfo {year} {2012})}\BibitemShut {NoStop}%
\bibitem [{\citenamefont {McLellan}\ \emph {et~al.}(2016)\citenamefont
  {McLellan}, \citenamefont {Myers}, \citenamefont {Kraemer}, \citenamefont
  {Ohno}, \citenamefont {Awschalom},\ and\ \citenamefont
  {Bleszynski~Jayich}}]{mclellan2016}%
  \BibitemOpen
  \bibfield  {author} {\bibinfo {author} {\bibfnamefont {C.~A.}\ \bibnamefont
  {McLellan}}, \bibinfo {author} {\bibfnamefont {B.~A.}\ \bibnamefont {Myers}},
  \bibinfo {author} {\bibfnamefont {S.}~\bibnamefont {Kraemer}}, \bibinfo
  {author} {\bibfnamefont {K.}~\bibnamefont {Ohno}}, \bibinfo {author}
  {\bibfnamefont {D.~D.}\ \bibnamefont {Awschalom}}, \ and\ \bibinfo {author}
  {\bibfnamefont {A.~C.}\ \bibnamefont {Bleszynski~Jayich}},\ }\href@noop {}
  {\bibfield  {journal} {\bibinfo  {journal} {Nano Letters}\ }\textbf {\bibinfo
  {volume} {16}},\ \bibinfo {pages} {2450} (\bibinfo {year}
  {2016})}\BibitemShut {NoStop}%
\bibitem [{\citenamefont {Eaton}\ \emph {et~al.}(2010)\citenamefont {Eaton},
  \citenamefont {Eaton}, \citenamefont {Barr},\ and\ \citenamefont
  {Weber}}]{Eaton2010}%
  \BibitemOpen
  \bibfield  {author} {\bibinfo {author} {\bibfnamefont {G.~R.}\ \bibnamefont
  {Eaton}}, \bibinfo {author} {\bibfnamefont {S.~S.}\ \bibnamefont {Eaton}},
  \bibinfo {author} {\bibfnamefont {D.~P.}\ \bibnamefont {Barr}}, \ and\
  \bibinfo {author} {\bibfnamefont {R.~T.}\ \bibnamefont {Weber}},\ }\href@noop
  {} {\emph {\bibinfo {title} {Quantitative Epr}}}\ (\bibinfo  {publisher}
  {Springer Science \& Business Media},\ \bibinfo {year} {2010})\BibitemShut
  {NoStop}%
\bibitem [{\citenamefont {Cox}\ \emph {et~al.}(1994)\citenamefont {Cox},
  \citenamefont {Newton},\ and\ \citenamefont {Baker}}]{cox199413c}%
  \BibitemOpen
  \bibfield  {author} {\bibinfo {author} {\bibfnamefont {A.}~\bibnamefont
  {Cox}}, \bibinfo {author} {\bibfnamefont {M.}~\bibnamefont {Newton}}, \ and\
  \bibinfo {author} {\bibfnamefont {J.}~\bibnamefont {Baker}},\ }\href@noop {}
  {\bibfield  {journal} {\bibinfo  {journal} {Journal of Physics: Condensed
  Matter}\ }\textbf {\bibinfo {volume} {6}},\ \bibinfo {pages} {551} (\bibinfo
  {year} {1994})}\BibitemShut {NoStop}%
\bibitem [{\citenamefont {Salikhov}\ \emph {et~al.}(1981)\citenamefont
  {Salikhov}, \citenamefont {Dzuba},\ and\ \citenamefont
  {Raitsimring}}]{salikhov1981theory}%
  \BibitemOpen
  \bibfield  {author} {\bibinfo {author} {\bibfnamefont {K.}~\bibnamefont
  {Salikhov}}, \bibinfo {author} {\bibfnamefont {S.-A.}\ \bibnamefont {Dzuba}},
  \ and\ \bibinfo {author} {\bibfnamefont {A.~M.}\ \bibnamefont
  {Raitsimring}},\ }\href@noop {} {\bibfield  {journal} {\bibinfo  {journal}
  {Journal of Magnetic Resonance (1969)}\ }\textbf {\bibinfo {volume} {42}},\
  \bibinfo {pages} {255} (\bibinfo {year} {1981})}\BibitemShut {NoStop}%
\bibitem [{\citenamefont {Romanelli}\ and\ \citenamefont
  {Kevan}(1997)}]{romanelli1997evaluation}%
  \BibitemOpen
  \bibfield  {author} {\bibinfo {author} {\bibfnamefont {M.}~\bibnamefont
  {Romanelli}}\ and\ \bibinfo {author} {\bibfnamefont {L.}~\bibnamefont
  {Kevan}},\ }\href@noop {} {\bibfield  {journal} {\bibinfo  {journal}
  {Concepts in Magnetic Resonance: An Educational Journal}\ }\textbf {\bibinfo
  {volume} {9}},\ \bibinfo {pages} {403} (\bibinfo {year} {1997})}\BibitemShut
  {NoStop}%
\bibitem [{\citenamefont {Stepanov}\ and\ \citenamefont
  {Takahashi}(2016)}]{stepanov2016determination}%
  \BibitemOpen
  \bibfield  {author} {\bibinfo {author} {\bibfnamefont {V.}~\bibnamefont
  {Stepanov}}\ and\ \bibinfo {author} {\bibfnamefont {S.}~\bibnamefont
  {Takahashi}},\ }\href@noop {} {\bibfield  {journal} {\bibinfo  {journal}
  {Physical Review B}\ }\textbf {\bibinfo {volume} {94}},\ \bibinfo {pages}
  {024421} (\bibinfo {year} {2016})}\BibitemShut {NoStop}%
\bibitem [{\citenamefont {Tyryshkin}\ \emph {et~al.}(2012)\citenamefont
  {Tyryshkin}, \citenamefont {Tojo}, \citenamefont {Morton}, \citenamefont
  {Riemann}, \citenamefont {Abrosimov}, \citenamefont {Becker}, \citenamefont
  {Pohl}, \citenamefont {Schenkel}, \citenamefont {Thewalt}, \citenamefont
  {Itoh} \emph {et~al.}}]{tyryshkin2012electron}%
  \BibitemOpen
  \bibfield  {author} {\bibinfo {author} {\bibfnamefont {A.~M.}\ \bibnamefont
  {Tyryshkin}}, \bibinfo {author} {\bibfnamefont {S.}~\bibnamefont {Tojo}},
  \bibinfo {author} {\bibfnamefont {J.~J.}\ \bibnamefont {Morton}}, \bibinfo
  {author} {\bibfnamefont {H.}~\bibnamefont {Riemann}}, \bibinfo {author}
  {\bibfnamefont {N.~V.}\ \bibnamefont {Abrosimov}}, \bibinfo {author}
  {\bibfnamefont {P.}~\bibnamefont {Becker}}, \bibinfo {author} {\bibfnamefont
  {H.-J.}\ \bibnamefont {Pohl}}, \bibinfo {author} {\bibfnamefont
  {T.}~\bibnamefont {Schenkel}}, \bibinfo {author} {\bibfnamefont {M.~L.}\
  \bibnamefont {Thewalt}}, \bibinfo {author} {\bibfnamefont {K.~M.}\
  \bibnamefont {Itoh}},  \emph {et~al.},\ }\href@noop {} {\bibfield  {journal}
  {\bibinfo  {journal} {Nature Materials}\ }\textbf {\bibinfo {volume} {11}},\
  \bibinfo {pages} {143} (\bibinfo {year} {2012})}\BibitemShut {NoStop}%
\bibitem [{\citenamefont {Wilson}\ \emph {et~al.}(2018)\citenamefont {Wilson},
  \citenamefont {Aronson}, \citenamefont {Clayton}, \citenamefont {Glaser},
  \citenamefont {Han},\ and\ \citenamefont {Sherwin}}]{wilson2018multi}%
  \BibitemOpen
  \bibfield  {author} {\bibinfo {author} {\bibfnamefont {C.~B.}\ \bibnamefont
  {Wilson}}, \bibinfo {author} {\bibfnamefont {S.}~\bibnamefont {Aronson}},
  \bibinfo {author} {\bibfnamefont {J.~A.}\ \bibnamefont {Clayton}}, \bibinfo
  {author} {\bibfnamefont {S.~J.}\ \bibnamefont {Glaser}}, \bibinfo {author}
  {\bibfnamefont {S.}~\bibnamefont {Han}}, \ and\ \bibinfo {author}
  {\bibfnamefont {M.}~\bibnamefont {Sherwin}},\ }\href@noop {} {\bibfield
  {journal} {\bibinfo  {journal} {Physical Chemistry Chemical Physics}\ }
  (\bibinfo {year} {2018})}\BibitemShut {NoStop}%
\bibitem [{\citenamefont {Rose}\ \emph {et~al.}(2017)\citenamefont {Rose},
  \citenamefont {Weis}, \citenamefont {Tyryshkin}, \citenamefont {Schenkel},\
  and\ \citenamefont {Lyon}}]{ROSE201732}%
  \BibitemOpen
  \bibfield  {author} {\bibinfo {author} {\bibfnamefont {B.}~\bibnamefont
  {Rose}}, \bibinfo {author} {\bibfnamefont {C.}~\bibnamefont {Weis}}, \bibinfo
  {author} {\bibfnamefont {A.}~\bibnamefont {Tyryshkin}}, \bibinfo {author}
  {\bibfnamefont {T.}~\bibnamefont {Schenkel}}, \ and\ \bibinfo {author}
  {\bibfnamefont {S.}~\bibnamefont {Lyon}},\ }\href {\doibase
  https://doi.org/10.1016/j.diamond.2016.12.009} {\bibfield  {journal}
  {\bibinfo  {journal} {Diamond and Related Materials}\ }\textbf {\bibinfo
  {volume} {72}},\ \bibinfo {pages} {32 } (\bibinfo {year} {2017})}\BibitemShut
  {NoStop}%
\bibitem [{\citenamefont {Kevan}\ and\ \citenamefont
  {Schwartz}(1979)}]{1979Tdes}%
  \BibitemOpen
  \bibfield  {author} {\bibinfo {author} {\bibfnamefont {L.}~\bibnamefont
  {Kevan}}\ and\ \bibinfo {author} {\bibfnamefont {R.~N.}\ \bibnamefont
  {Schwartz}},\ }\href@noop {} {\emph {\bibinfo {title} {Time domain electron
  spin resonance}}}\ (\bibinfo  {publisher} {Wiley},\ \bibinfo {address} {New
  York},\ \bibinfo {year} {1979})\BibitemShut {NoStop}%
\bibitem [{\citenamefont {Klauder}\ and\ \citenamefont
  {Anderson}(1962)}]{klauder1962spectral}%
  \BibitemOpen
  \bibfield  {author} {\bibinfo {author} {\bibfnamefont {J.}~\bibnamefont
  {Klauder}}\ and\ \bibinfo {author} {\bibfnamefont {P.}~\bibnamefont
  {Anderson}},\ }\href@noop {} {\bibfield  {journal} {\bibinfo  {journal}
  {Physical Review}\ }\textbf {\bibinfo {volume} {125}},\ \bibinfo {pages}
  {912} (\bibinfo {year} {1962})}\BibitemShut {NoStop}%
\bibitem [{\citenamefont {Milov}\ \emph {et~al.}(1998)\citenamefont {Milov},
  \citenamefont {Maryasov},\ and\ \citenamefont {Tsvetkov}}]{Milov1998}%
  \BibitemOpen
  \bibfield  {author} {\bibinfo {author} {\bibfnamefont {A.~D.}\ \bibnamefont
  {Milov}}, \bibinfo {author} {\bibfnamefont {A.~G.}\ \bibnamefont {Maryasov}},
  \ and\ \bibinfo {author} {\bibfnamefont {Y.~D.}\ \bibnamefont {Tsvetkov}},\
  }\href {\doibase 10.1007/BF03161886} {\bibfield  {journal} {\bibinfo
  {journal} {Applied Magnetic Resonance}\ }\textbf {\bibinfo {volume} {15}},\
  \bibinfo {pages} {107} (\bibinfo {year} {1998})}\BibitemShut {NoStop}%
\bibitem [{\citenamefont {Rowan}\ \emph {et~al.}(1965)\citenamefont {Rowan},
  \citenamefont {Hahn},\ and\ \citenamefont {Mims}}]{rowan1965electron}%
  \BibitemOpen
  \bibfield  {author} {\bibinfo {author} {\bibfnamefont {L.}~\bibnamefont
  {Rowan}}, \bibinfo {author} {\bibfnamefont {E.}~\bibnamefont {Hahn}}, \ and\
  \bibinfo {author} {\bibfnamefont {W.}~\bibnamefont {Mims}},\ }\href@noop {}
  {\bibfield  {journal} {\bibinfo  {journal} {Physical Review}\ }\textbf
  {\bibinfo {volume} {137}},\ \bibinfo {pages} {A61} (\bibinfo {year}
  {1965})}\BibitemShut {NoStop}%
\bibitem [{\citenamefont {Campbell}\ and\ \citenamefont
  {Mainwood}(2000)}]{campbell2000radiation}%
  \BibitemOpen
  \bibfield  {author} {\bibinfo {author} {\bibfnamefont {B.}~\bibnamefont
  {Campbell}}\ and\ \bibinfo {author} {\bibfnamefont {A.}~\bibnamefont
  {Mainwood}},\ }\href@noop {} {\bibfield  {journal} {\bibinfo  {journal}
  {Physica Status Solidi (A)}\ }\textbf {\bibinfo {volume} {181}},\ \bibinfo
  {pages} {99} (\bibinfo {year} {2000})}\BibitemShut {NoStop}%
\bibitem [{\citenamefont {Koike}\ \emph {et~al.}(1992)\citenamefont {Koike},
  \citenamefont {Parkin},\ and\ \citenamefont
  {Mitchell}}]{koike1992displacement}%
  \BibitemOpen
  \bibfield  {author} {\bibinfo {author} {\bibfnamefont {J.}~\bibnamefont
  {Koike}}, \bibinfo {author} {\bibfnamefont {D.}~\bibnamefont {Parkin}}, \
  and\ \bibinfo {author} {\bibfnamefont {T.}~\bibnamefont {Mitchell}},\
  }\href@noop {} {\bibfield  {journal} {\bibinfo  {journal} {Applied Physics
  Letters}\ }\textbf {\bibinfo {volume} {60}},\ \bibinfo {pages} {1450}
  (\bibinfo {year} {1992})}\BibitemShut {NoStop}%
\bibitem [{\citenamefont {Clark}\ \emph {et~al.}(1961)\citenamefont {Clark},
  \citenamefont {Kemmey},\ and\ \citenamefont {Mitchell}}]{clark1961optical}%
  \BibitemOpen
  \bibfield  {author} {\bibinfo {author} {\bibfnamefont {C.}~\bibnamefont
  {Clark}}, \bibinfo {author} {\bibfnamefont {P.}~\bibnamefont {Kemmey}}, \
  and\ \bibinfo {author} {\bibfnamefont {E.}~\bibnamefont {Mitchell}},\
  }\href@noop {} {\bibfield  {journal} {\bibinfo  {journal} {Discussions of the
  Faraday Society}\ }\textbf {\bibinfo {volume} {31}},\ \bibinfo {pages} {96}
  (\bibinfo {year} {1961})}\BibitemShut {NoStop}%
\bibitem [{\citenamefont {Wang}\ and\ \citenamefont
  {Takahashi}(2013)}]{wang2013spin}%
  \BibitemOpen
  \bibfield  {author} {\bibinfo {author} {\bibfnamefont {Z.-H.}\ \bibnamefont
  {Wang}}\ and\ \bibinfo {author} {\bibfnamefont {S.}~\bibnamefont
  {Takahashi}},\ }\href@noop {} {\bibfield  {journal} {\bibinfo  {journal}
  {Physical Review B}\ }\textbf {\bibinfo {volume} {87}},\ \bibinfo {pages}
  {115122} (\bibinfo {year} {2013})}\BibitemShut {NoStop}%
\bibitem [{\citenamefont {Pham}(2013)}]{pham2013magnetic}%
  \BibitemOpen
  \bibfield  {author} {\bibinfo {author} {\bibfnamefont {L.~M.}\ \bibnamefont
  {Pham}},\ }\emph {\bibinfo {title} {Magnetic Field Sensing with
  Nitrogen-Vacancy Color Centers in Diamond}},\ \href@noop {} {Ph.D. thesis},\
  \bibinfo  {school} {Harvard University} (\bibinfo {year} {2013})\BibitemShut
  {NoStop}%
\bibitem [{\citenamefont {Schloss}\ \emph {et~al.}(2018)\citenamefont
  {Schloss}, \citenamefont {Barry}, \citenamefont {Turner},\ and\ \citenamefont
  {Walsworth}}]{schloss2018simultaneous}%
  \BibitemOpen
  \bibfield  {author} {\bibinfo {author} {\bibfnamefont {J.~M.}\ \bibnamefont
  {Schloss}}, \bibinfo {author} {\bibfnamefont {J.~F.}\ \bibnamefont {Barry}},
  \bibinfo {author} {\bibfnamefont {M.~J.}\ \bibnamefont {Turner}}, \ and\
  \bibinfo {author} {\bibfnamefont {R.~L.}\ \bibnamefont {Walsworth}},\
  }\href@noop {} {\bibfield  {journal} {\bibinfo  {journal} {arXiv preprint
  arXiv:1803.03718}\ } (\bibinfo {year} {2018})}\BibitemShut {NoStop}%
\bibitem [{\citenamefont {Chu}\ \emph {et~al.}(2014)\citenamefont {Chu},
  \citenamefont {de~Leon}, \citenamefont {Shields}, \citenamefont {Hausmann},
  \citenamefont {Evans}, \citenamefont {Togan}, \citenamefont {Burek},
  \citenamefont {Markham}, \citenamefont {Stacey}, \citenamefont {Zibrov} \emph
  {et~al.}}]{chu2014coherent}%
  \BibitemOpen
  \bibfield  {author} {\bibinfo {author} {\bibfnamefont {Y.}~\bibnamefont
  {Chu}}, \bibinfo {author} {\bibfnamefont {N.~P.}\ \bibnamefont {de~Leon}},
  \bibinfo {author} {\bibfnamefont {B.~J.}\ \bibnamefont {Shields}}, \bibinfo
  {author} {\bibfnamefont {B.}~\bibnamefont {Hausmann}}, \bibinfo {author}
  {\bibfnamefont {R.}~\bibnamefont {Evans}}, \bibinfo {author} {\bibfnamefont
  {E.}~\bibnamefont {Togan}}, \bibinfo {author} {\bibfnamefont {M.~J.}\
  \bibnamefont {Burek}}, \bibinfo {author} {\bibfnamefont {M.}~\bibnamefont
  {Markham}}, \bibinfo {author} {\bibfnamefont {A.}~\bibnamefont {Stacey}},
  \bibinfo {author} {\bibfnamefont {A.~S.}\ \bibnamefont {Zibrov}},  \emph
  {et~al.},\ }\href@noop {} {\bibfield  {journal} {\bibinfo  {journal} {Nano
  Letters}\ }\textbf {\bibinfo {volume} {14}},\ \bibinfo {pages} {1982}
  (\bibinfo {year} {2014})}\BibitemShut {NoStop}%
\bibitem [{\citenamefont {Choi}\ \emph
  {et~al.}(2017{\natexlab{b}})\citenamefont {Choi}, \citenamefont {Choi},
  \citenamefont {Kucsko}, \citenamefont {Maurer}, \citenamefont {Shields},
  \citenamefont {Sumiya}, \citenamefont {Onoda}, \citenamefont {Isoya},
  \citenamefont {Demler}, \citenamefont {Jelezko} \emph
  {et~al.}}]{choi2017depolarization}%
  \BibitemOpen
  \bibfield  {author} {\bibinfo {author} {\bibfnamefont {J.}~\bibnamefont
  {Choi}}, \bibinfo {author} {\bibfnamefont {S.}~\bibnamefont {Choi}}, \bibinfo
  {author} {\bibfnamefont {G.}~\bibnamefont {Kucsko}}, \bibinfo {author}
  {\bibfnamefont {P.~C.}\ \bibnamefont {Maurer}}, \bibinfo {author}
  {\bibfnamefont {B.~J.}\ \bibnamefont {Shields}}, \bibinfo {author}
  {\bibfnamefont {H.}~\bibnamefont {Sumiya}}, \bibinfo {author} {\bibfnamefont
  {S.}~\bibnamefont {Onoda}}, \bibinfo {author} {\bibfnamefont
  {J.}~\bibnamefont {Isoya}}, \bibinfo {author} {\bibfnamefont
  {E.}~\bibnamefont {Demler}}, \bibinfo {author} {\bibfnamefont
  {F.}~\bibnamefont {Jelezko}},  \emph {et~al.},\ }\href@noop {} {\bibfield
  {journal} {\bibinfo  {journal} {Physical Review Letters}\ }\textbf {\bibinfo
  {volume} {118}},\ \bibinfo {pages} {093601} (\bibinfo {year}
  {2017}{\natexlab{b}})}\BibitemShut {NoStop}%
\bibitem [{\citenamefont {Kucsko}\ \emph {et~al.}(2018)\citenamefont {Kucsko},
  \citenamefont {Choi}, \citenamefont {Choi}, \citenamefont {Maurer},
  \citenamefont {Zhou}, \citenamefont {Landig}, \citenamefont {Sumiya},
  \citenamefont {Onoda}, \citenamefont {Isoya}, \citenamefont {Jelezko},
  \citenamefont {Demler}, \citenamefont {Yao},\ and\ \citenamefont
  {Lukin}}]{kucsko2018critical}%
  \BibitemOpen
  \bibfield  {author} {\bibinfo {author} {\bibfnamefont {G.}~\bibnamefont
  {Kucsko}}, \bibinfo {author} {\bibfnamefont {S.}~\bibnamefont {Choi}},
  \bibinfo {author} {\bibfnamefont {J.}~\bibnamefont {Choi}}, \bibinfo {author}
  {\bibfnamefont {P.~C.}\ \bibnamefont {Maurer}}, \bibinfo {author}
  {\bibfnamefont {H.}~\bibnamefont {Zhou}}, \bibinfo {author} {\bibfnamefont
  {R.}~\bibnamefont {Landig}}, \bibinfo {author} {\bibfnamefont
  {H.}~\bibnamefont {Sumiya}}, \bibinfo {author} {\bibfnamefont
  {S.}~\bibnamefont {Onoda}}, \bibinfo {author} {\bibfnamefont
  {J.}~\bibnamefont {Isoya}}, \bibinfo {author} {\bibfnamefont
  {F.}~\bibnamefont {Jelezko}}, \bibinfo {author} {\bibfnamefont
  {E.}~\bibnamefont {Demler}}, \bibinfo {author} {\bibfnamefont {N.~Y.}\
  \bibnamefont {Yao}}, \ and\ \bibinfo {author} {\bibfnamefont {M.~D.}\
  \bibnamefont {Lukin}},\ }\href {\doibase 10.1103/PhysRevLett.121.023601}
  {\bibfield  {journal} {\bibinfo  {journal} {Physical Review Letters}\
  }\textbf {\bibinfo {volume} {121}},\ \bibinfo {pages} {023601} (\bibinfo
  {year} {2018})}\BibitemShut {NoStop}%
\bibitem [{\citenamefont {Yao}\ \emph {et~al.}(2014)\citenamefont {Yao},
  \citenamefont {Laumann}, \citenamefont {Gopalakrishnan}, \citenamefont
  {Knap}, \citenamefont {Mueller}, \citenamefont {Demler},\ and\ \citenamefont
  {Lukin}}]{yao2014many}%
  \BibitemOpen
  \bibfield  {author} {\bibinfo {author} {\bibfnamefont {N.~Y.}\ \bibnamefont
  {Yao}}, \bibinfo {author} {\bibfnamefont {C.~R.}\ \bibnamefont {Laumann}},
  \bibinfo {author} {\bibfnamefont {S.}~\bibnamefont {Gopalakrishnan}},
  \bibinfo {author} {\bibfnamefont {M.}~\bibnamefont {Knap}}, \bibinfo {author}
  {\bibfnamefont {M.}~\bibnamefont {Mueller}}, \bibinfo {author} {\bibfnamefont
  {E.~A.}\ \bibnamefont {Demler}}, \ and\ \bibinfo {author} {\bibfnamefont
  {M.~D.}\ \bibnamefont {Lukin}},\ }\href@noop {} {\bibfield  {journal}
  {\bibinfo  {journal} {Physical Review Letters}\ }\textbf {\bibinfo {volume}
  {113}},\ \bibinfo {pages} {243002} (\bibinfo {year} {2014})}\BibitemShut
  {NoStop}%
\bibitem [{\citenamefont {Choi}\ \emph
  {et~al.}(2017{\natexlab{c}})\citenamefont {Choi}, \citenamefont {Choi},
  \citenamefont {Landig}, \citenamefont {Kucsko}, \citenamefont {Zhou},
  \citenamefont {Isoya}, \citenamefont {Jelezko}, \citenamefont {Onoda},
  \citenamefont {Sumiya}, \citenamefont {Khemani} \emph
  {et~al.}}]{choi2017observation}%
  \BibitemOpen
  \bibfield  {author} {\bibinfo {author} {\bibfnamefont {S.}~\bibnamefont
  {Choi}}, \bibinfo {author} {\bibfnamefont {J.}~\bibnamefont {Choi}}, \bibinfo
  {author} {\bibfnamefont {R.}~\bibnamefont {Landig}}, \bibinfo {author}
  {\bibfnamefont {G.}~\bibnamefont {Kucsko}}, \bibinfo {author} {\bibfnamefont
  {H.}~\bibnamefont {Zhou}}, \bibinfo {author} {\bibfnamefont {J.}~\bibnamefont
  {Isoya}}, \bibinfo {author} {\bibfnamefont {F.}~\bibnamefont {Jelezko}},
  \bibinfo {author} {\bibfnamefont {S.}~\bibnamefont {Onoda}}, \bibinfo
  {author} {\bibfnamefont {H.}~\bibnamefont {Sumiya}}, \bibinfo {author}
  {\bibfnamefont {V.}~\bibnamefont {Khemani}},  \emph {et~al.},\ }\href@noop {}
  {\bibfield  {journal} {\bibinfo  {journal} {Nature}\ }\textbf {\bibinfo
  {volume} {543}},\ \bibinfo {pages} {221} (\bibinfo {year}
  {2017}{\natexlab{c}})}\BibitemShut {NoStop}%
\bibitem [{\citenamefont {Rovny}\ \emph {et~al.}(2018)\citenamefont {Rovny},
  \citenamefont {Blum},\ and\ \citenamefont {Barrett}}]{rovny2018observation}%
  \BibitemOpen
  \bibfield  {author} {\bibinfo {author} {\bibfnamefont {J.}~\bibnamefont
  {Rovny}}, \bibinfo {author} {\bibfnamefont {R.~L.}\ \bibnamefont {Blum}}, \
  and\ \bibinfo {author} {\bibfnamefont {S.~E.}\ \bibnamefont {Barrett}},\
  }\href@noop {} {\bibfield  {journal} {\bibinfo  {journal} {Physical Review
  Letters}\ }\textbf {\bibinfo {volume} {120}},\ \bibinfo {pages} {180603}
  (\bibinfo {year} {2018})}\BibitemShut {NoStop}%
\bibitem [{\citenamefont {Putz}\ \emph {et~al.}(2014)\citenamefont {Putz},
  \citenamefont {Krimer}, \citenamefont {Amsuess}, \citenamefont {Valookaran},
  \citenamefont {Noebauer}, \citenamefont {Schmiedmayer}, \citenamefont
  {Rotter},\ and\ \citenamefont {Majer}}]{putz2014protecting}%
  \BibitemOpen
  \bibfield  {author} {\bibinfo {author} {\bibfnamefont {S.}~\bibnamefont
  {Putz}}, \bibinfo {author} {\bibfnamefont {D.~O.}\ \bibnamefont {Krimer}},
  \bibinfo {author} {\bibfnamefont {R.}~\bibnamefont {Amsuess}}, \bibinfo
  {author} {\bibfnamefont {A.}~\bibnamefont {Valookaran}}, \bibinfo {author}
  {\bibfnamefont {T.}~\bibnamefont {Noebauer}}, \bibinfo {author}
  {\bibfnamefont {J.}~\bibnamefont {Schmiedmayer}}, \bibinfo {author}
  {\bibfnamefont {S.}~\bibnamefont {Rotter}}, \ and\ \bibinfo {author}
  {\bibfnamefont {J.}~\bibnamefont {Majer}},\ }\href@noop {} {\bibfield
  {journal} {\bibinfo  {journal} {Nature Physics}\ }\textbf {\bibinfo {volume}
  {10}},\ \bibinfo {pages} {720} (\bibinfo {year} {2014})}\BibitemShut
  {NoStop}%
\end{thebibliography}%

\end{document}